\documentclass[twocolumn,prl,amsmath,amssymb, aps,superscriptaddress, nofootinbib]{revtex4-2}
 
\usepackage[titletoc,toc,title]{appendix}
\usepackage{titletoc}
\usepackage{titlesec}
\usepackage[version=4]{mhchem} 
\usepackage[normalem]{ulem}
\usepackage{mhchem}
\usepackage{amsmath}
\usepackage{amssymb}
\usepackage{amsfonts}
\usepackage{graphicx}
\usepackage{xcolor}
\usepackage{xfrac}
\usepackage{footmisc}
\usepackage{comment}
\usepackage{pifont}
\usepackage{times}
\usepackage[hidelinks]{hyperref}
\usepackage{enumitem}
\usepackage{centernot}
\usepackage{cancel}
\usepackage{slashed}
\usepackage{relsize}
\usepackage{amsmath,amssymb,mathrsfs}
\usepackage{times}
\usepackage{epsfig}
\usepackage{verbatim}
\usepackage{bm}
\usepackage[utf8]{inputenc}
\usepackage{graphics}
\usepackage{graphicx,epsfig,amssymb,amsmath,color,cancel}
\usepackage{subfigure}
\usepackage[english]{babel}
\usepackage{adjustbox}
\usepackage{physics}

\usepackage{array}
\newcolumntype{P}[1]{>{\centering\arraybackslash}p{#1}}
\newcolumntype{M}[1]{>{\centering\arraybackslash}m{#1}}
\usepackage{tabularray}

\usepackage{orcidlink}
\usepackage{tabularray}
\UseTblrLibrary{booktabs}

\definecolor{zima_blue}{HTML}{1393C1}
\hypersetup{setpagesize=false,bookmarksnumbered=true,bookmarksopen=true,%
colorlinks=true,linkcolor=zima_blue,urlcolor=zima_blue,citecolor=zima_blue,linktocpage=false}

\begin{document}

\preprint{}

\title{First-Order Phase Transition Interpretation of Pulsar Timing Array Signal \\
Produces Solar-Mass Black Holes}

\author{Yann Gouttenoire~\orcidlink{0000-0003-2225-6704}}
\email{yann.gouttenoire@gmail.com}
\affiliation{School of Physics and Astronomy, Tel-Aviv University, Tel-Aviv 69978, Israel}

\begin{abstract}
 
We perform a Bayesian analysis of NANOGrav 15yr and IPTA DR2 pulsar timing residuals and show that the recently detected stochastic gravitational-wave background (SGWB) is compatible with a SGWB produced by bubble dynamics during a cosmological first-order phase transition. The timing data suggests that the phase transition would occur around QCD confinement temperature and would have a slow rate of completion. This scenario can naturally lead to the abundant production of primordial black holes (PBHs) with solar masses. These PBHs can potentially be detected by current and advanced gravitational wave detectors LIGO-Virgo-Kagra, Einstein Telescope, Cosmic Explorer, by astrometry with GAIA and by 21-cm survey.

\vspace{0.4cm}
\noindent
DOI:~\href{https://doi.org/10.1103/PhysRevLett.131.171404}{ 10.1103/PhysRevLett.131.171404}

\end{abstract}

\maketitle

\section{INTRODUCTION}

\begin{figure*}[ht!]
    \centering
    \includegraphics[width=0.7\textwidth]{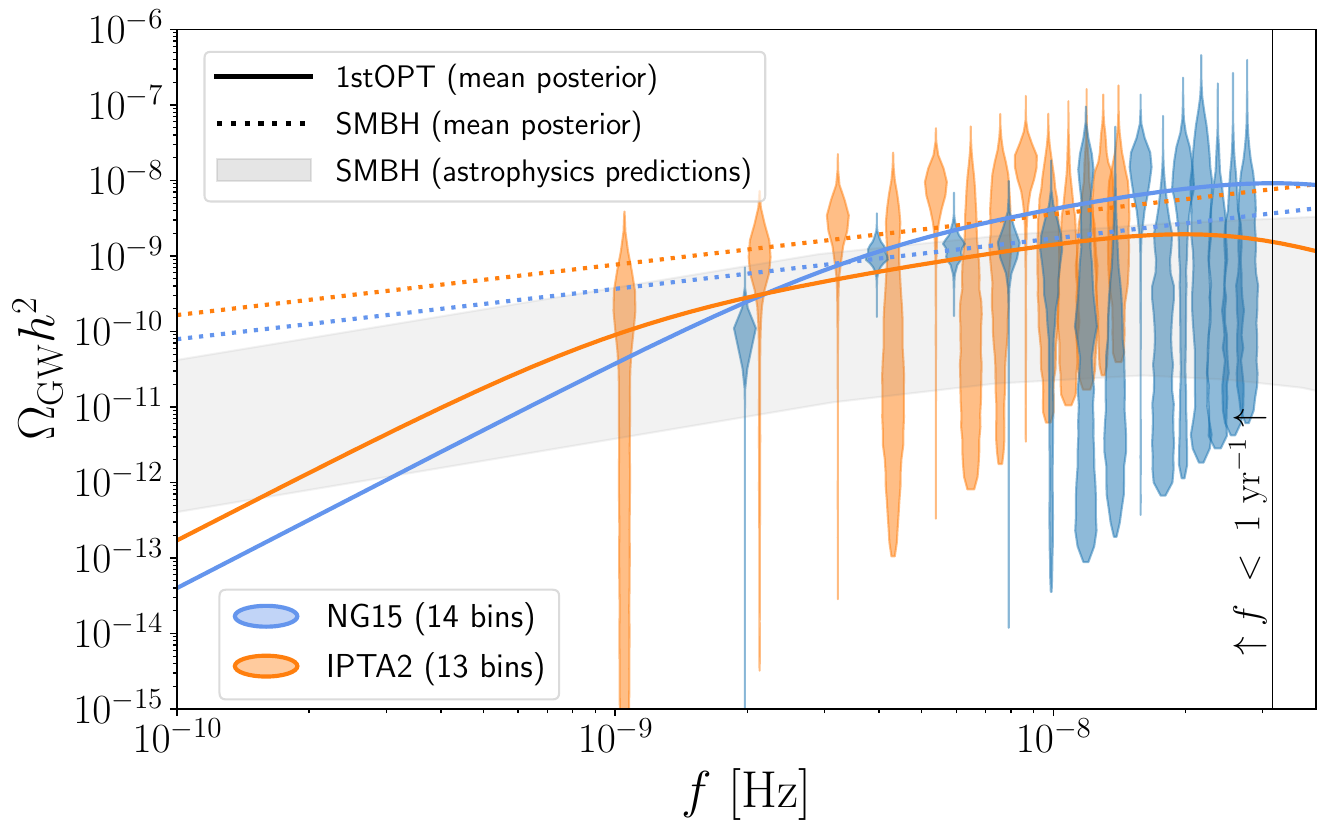}
      \caption{\label{fig:1stOPT_trapeze} \small  \small    The \textbf{violin} diagrams depict the posterior probability distribution of the SGWB energy density in each frequency bins of \href{https://zenodo.org/record/8060824}{NG15} and
\href{https://zenodo.org/record/5787557}{IPTA2} data sets. We overlay with \textbf{solid} lines the SWGB from 1stOPT, obtained using Eq.~\eqref{eq:Bulk_flow}, using mean posterior value for the PT parameters. The \textbf{dotted} lines illustrate the SGWB originating from SMBH binaries, employing the mean posterior value for the amplitude and fixing the power-law index to $\beta = 2/3$. The \textbf{gray} band represents the 90$\%$ confidence interval for the projected SGWB based on a Monte Carlo simulation of a binary population of SMBHs \cite{Rosado:2015epa}.}
\end{figure*}

By measuring cross-correlations in the arrival times of pulses emitted by rotating neutron stars, Pulsar Timing Arrays (PTAs) have been established as a mean to detect nano-Hertz (nHz) frequency Gravitational Waves (GW). 
In 2020, a common low-frequency noise has been identified in the datasets of NANOGrav \cite{NANOGrav:2020spf}, EPTA \cite{Chen:2021rqp}, and PPTA \cite{Goncharov:2021oub}, and confirmed in 2022 by IPTA \cite{Antoniadis:2022pcn} (IPTA2) which combines data from the former.
To distinguish a GW origin from systematic effects requires timing delay correlations to have a quadrupolar dependence on the angular separation between pulsars \cite{Hellings:1983fr}. In June 2023, following the analysis of their most recent data, the collaborative efforts of NANOGrav, EPTA and PPTA (NG15, EPTA2 and PPTA3) have identified compelling statistical evidence for such interpulsar correlations \cite{NANOGrav:2023gor,Antoniadis:2023ott,Reardon:2023gzh}, with Bayes factors of 600, 60, and 11, respectively.
The primary expected source of GWs at low frequencies is believed to be from supermassive black holes binaries (SMBH) \cite{Sesana:2013wja,Kelley:2017lek,Chen:2018znx}. The stochastic GW background (SGWB) inferred from PTA data corresponds to the upper limit of the astrophysical predicted interval, see Fig.~\ref{fig:1stOPT_trapeze}. This could suggest that SMBH binaries are slightly more massive and more numerous than initially anticipated \cite{Middleton:2020asl,Casey-Clyde:2021xro,NANOGrav:2023hfp,Antoniadis:2023xlr}.  Alternatively, the PTA SGWB might originate from new physics taking place in the early universe \cite{NANOGrav:2023hvm,Antoniadis:2023xlr,Madge:2023cak,Figueroa:2023zhu}. The last hypothesis however comes with its own set of challenges.  For instance, ascribing the SGWB to inflation necessitates unnaturally large values for the spectral tilt $n_t \simeq 1.8$ and a low reheating temperature $T_{\rm reh} \lesssim 10~\rm GeV$ \cite{Vagnozzi:2023lwo}. GW induced by a Gaussian spectrum of curvature perturbation would results in excessive PBHs production \cite{Chen:2019xse,Dandoy:2023jot,Franciolini:2023pbf}. A SGWB resulting from PBH mergers would not align with structure formation \cite{Gouttenoire:2023nzr,Depta:2023qst}. A cosmic strings network, when arising from a global symmetry is excluded by  Big-Bang Nucleosynthesis (BBN) \cite{Gorghetto:2021fsn,Chang:2021afa,Dror:2021nyr,Servant:2023mwt}, while when arising from a local symmetry is not favoured by the Bayesian analysis \cite{NANOGrav:2023hvm,Ellis:2023tsl}. 
To evade BBN bound, a first-order phase transition (1stOPT) sourcing PTA signal would necessitate the latent heat to be released dominantly to the Standard Model (SM), e.g. \cite{Ratzinger:2020koh,NANOGrav:2021flc,Bai:2021ibt,Bringmann:2023opz,NANOGrav:2023hvm,Antoniadis:2023xlr,Madge:2023cak,Figueroa:2023zhu}.
Interestingly however, the 1stOPT interpretation of PTA SGWB requires a reheating temperature around the scale of QCD confinement $100$ $\rm MeV$, with a rather low completion rate $\beta/H \lesssim 12$ and a large latent heat fraction $\alpha \gtrsim 0.5$ \cite{NANOGrav:2023hvm}.
This overlaps with the region where 1stOPT have been recently found to produce PBHs in observable amount \cite{Gouttenoire:2023naa}. The PBH prior has been omitted in all previous analysis of the 1stOPT interpretation of PTA data \cite{Ratzinger:2020koh,NANOGrav:2021flc,Bai:2021ibt,Bringmann:2023opz,NANOGrav:2023hvm,Antoniadis:2023xlr,Madge:2023cak,Figueroa:2023zhu,Nakai:2020oit,Addazi:2020zcj,Moore:2021ibq,Li:2021qer,Brandenburg:2021tmp,RoperPol:2022iel,Fujikura:2023lkn,Addazi:2023jvg,Xiao:2023dbb,Ghosh:2023aum,Yang:2023qlf,Athron:2023mer}. Similarly, domain wall networks annihilating into SM degrees of freedom are credible early-universe interpretation of the PTA signal associated with the production of of multi-solar-mass PBHs \cite{Gouttenoire:2023ftk,Gouttenoire:2023gbn}.
 
In this \textit{letter}, we perform a Bayesian search for SGWB from 1stOPT in NANOGrav 15-year (NG15) and IPTA DR2 (IPTA2) timing residuals, including both BBN-$N_{\rm eff}$-bound and PBH-overproduction constraints as priors in the analysis. 
To simplify the numerical strategy, we focus on the region $\alpha \gg 1$ of strong supercooling, where PBH production is the most efficient. In this region, the dependency of the GW signal on both the wall velocity $(v_w=1)$ and the latent heat fraction $\alpha$ disappears.\footnote{The Bayesian analysis of 1stOPT with finite $\alpha$ will be presented elsewhere.}
We argue that the SGWB from 1stOPT is given by the bulk flow model independently of whether the latent heat is still stored in bubble walls at percolation or has been released to the plasma before.
 We find that PBH formation does not exclude the 1stOPT interpretation of PTA signal. Instead, a SGWB from supercooled PT is favoured with respect to the SMBH binary hypothesis by a Bayes factor of $15$ in NG15 data set. 
We point for the first time, the existence of a multi-messenger window: the NG15 posterior contains a region producing  $[0.1-10]$ solar-mass PBHs, see Fig.~\ref{fig:beta_vs_Treh_allconstraints}. The merging of such PBHs could source GWs with kHz frequencies in the range of LIGO-Virgo
\cite{Nakamura:1997sm,Raidal:2018bbj,Kavanagh:2018ggo,LIGOScientific:2019kan,DeLuca:2020qqa}, and ET/CE \cite{Chen:2019irf,Pujolas:2021yaw}. Additionally, their presence could be detected from lensing in GAIA \cite{Chen:2023xyj,VanTilburg:2018ykj,Verma:2022pym} or from heating in 21-cm survey \cite{Mena:2019nhm,Villanueva-Domingo:2021cgh,Villanueva-Domingo:2021spv}.

We also consider the negative hypothesis in which the SGWB observed in PTA would not result from a supercooled PT and derive lower limits on the rate of completion $\beta/H \gtrsim [10-20]$, implying that the universe could not have boiled longer than $[5\%-10\%]$ of a Hubble time during the QCD phase transition.

\section{GRAVITATIONAL WAVES FROM FIRST-ORDER PT}
\underline{PT parameters} ---
The strength of a 1stOPT is characterized by the ratio of its latent heat $\Delta V$, defined as the vacuum energy difference between the two minima of the potential driving the transition, to the radiation energy density $\rho_{\rm rad}(T_n)$ at the nucleation temperature $T_n$
\begin{equation}
\label{eq:alpha_def}
    \alpha \equiv \frac{\Delta V}{\rho_{\rm rad}(T_n)} \equiv \left( \frac{T_{\rm eq}}{T_{\rm n}} \right)^4,
\end{equation}
where we have neglected a ratio of number of relativistic degrees of freedom.
A 1stOPT is said supercooled when $\alpha \gtrsim 1$, in which case the universe enters a stage of vacuum-domination at temperature $T_{\rm eq}$ which ends at $T_n$ when bubble growth converts the latent heat into radiation energy density. 
The rate at which nucleation takes place is controlled by the time derivative of the tunneling rate per unit of volume $\Gamma_{\rm \mathsmaller{V}}$
\begin{equation}
\label{eq:beta_def}
    \beta \equiv \frac{1}{\Gamma_{\rm \mathsmaller{V}}}\frac{d\Gamma_{\rm \mathsmaller{V}}}{dt}.
\end{equation} 
After the phase transition completes, the universe is reheated back to the temperature $T_{\rm eq}$ up to changes in number of degrees of freedom which we again neglect. 

\underline{Energy budget} ---
The dynamics of weak phase transition $\alpha < 1$ is rather well understood \cite{Caprini:2015zlo,Caprini:2019egz}. The non-relativistic motion of bubble walls, $\gamma_w\simeq 1$, converts the latent heat into thermal and kinetic energy of the plasma, which propagate under the form of long-lasting sound waves \cite{Espinosa:2010hh}, and ultimately turn into turbulence \cite{Gogoberidze:2007an,Caprini:2009yp,RoperPol:2019wvy,Niksa:2018ofa,Auclair:2022jod}. GWs sourced by sound waves have been intensively simulated on the lattice in the recent years \cite{Hindmarsh:2013xza, Hindmarsh:2015qta, Hindmarsh:2017gnf,Jinno:2020eqg,Jinno:2022mie}, and analytical modelling have been proposed \cite{Hindmarsh:2016lnk,Hindmarsh:2019phv}.
The dynamics of supercooled phase transition $\alpha >1$ is more complex due to the large Lorentz factor $\gamma_w\gg 1$ of bubble walls \cite{Bodeker:2017cim,Gouttenoire:2021kjv}.
In the relativistic limit, the acceleration of bubble walls with tension $\sigma$ is set by the pressure balance \cite{Gouttenoire:2023naa} 
\begin{equation}
    \frac{d\gamma_w}{dt}  = \frac{\Delta V-\mathcal{P}_{\rm fric}}{\sigma}.
\end{equation}
The friction pressure $\mathcal{P}_{\rm fric}$ is dominantly induced by transition radiation \cite{Bodeker:2017cim}, which resummed at leading-logs, reads \cite{Gouttenoire:2021kjv}
\begin{equation}
\mathcal{P}_{\rm fric} = c_0\,g_{\rm D}^3 \gamma_{w} v_{\phi} T_{n}^3 \log\left( \frac{v_{\phi}}{T_n}\right),\qquad c_0=\mathcal{O}(1),
\end{equation}
where $g_{\rm D}$ is a gauge coupling and $v_{\phi}$ is the vev of the scalar field driving the phase transition.
As bubble walls accelerate, the retarding pressure $\mathcal{P}_{\rm fric}$ grows linearly with $\gamma_w$.

\begin{figure*}[t]
    \centering
    \includegraphics[width=0.49\textwidth]{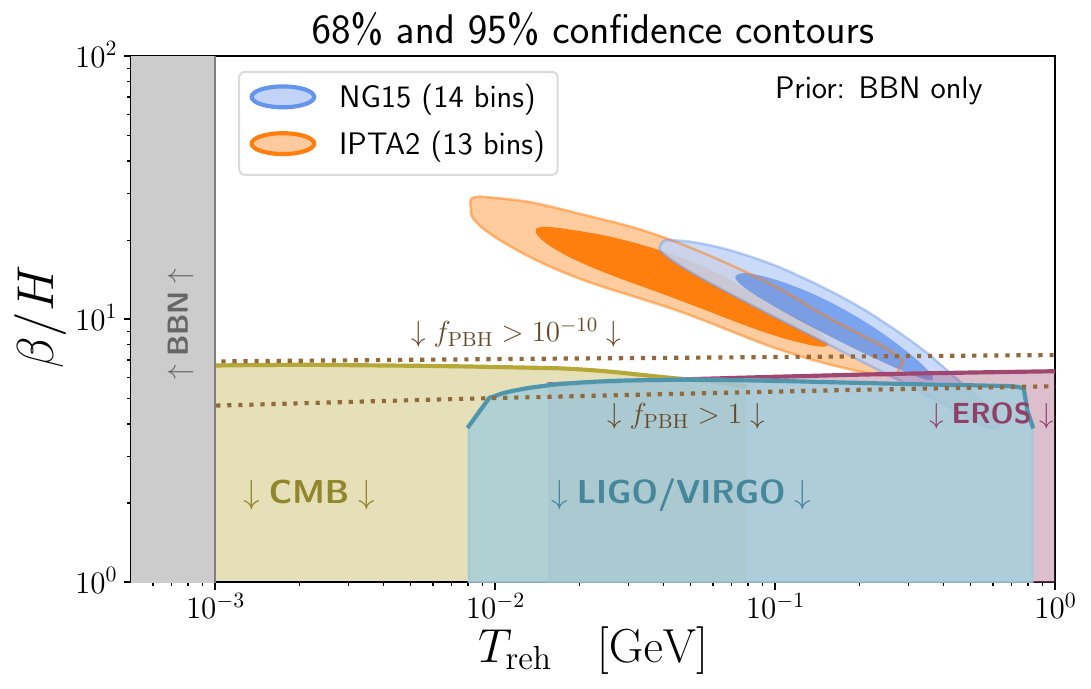}
    \includegraphics[width=0.47\textwidth]{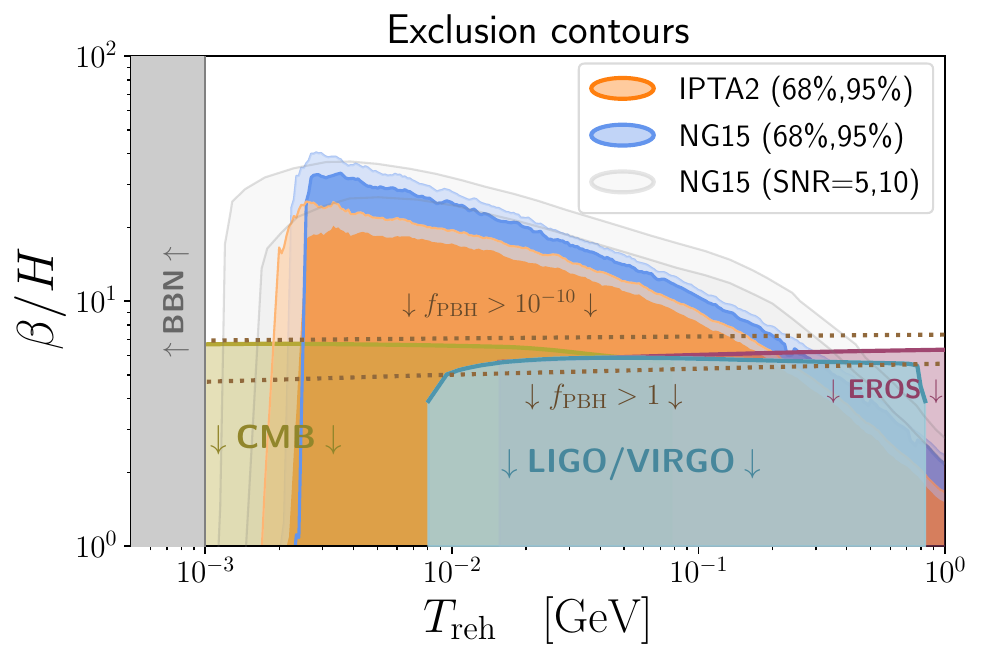}
      \caption{\label{fig:betaOH_TnOTeq} \small  \small     \textbf{Left}: Colored regions are posterior distributions in term of the reheating temperature $T_{\rm reh}$ and rate of completion $\beta/H$ of a strong 1stOPT ($\alpha\gg 1$). They are obtained after performing a Bayesian analysis of PTA dataset.  We overlay the CMB, LIGO/Virgo and microlensing (EROS) constraints on PBHs produced during such 1stOPT.  \textbf{Right}: Lower limit on the rate of completion $\beta/H$ in the negative hypothesis in which the PTA SGWB would not arise from a strong PT ($\alpha\gg 1$). We cast $68\%$ and $95\%$ lower limit using Bayesian inference as explained in App.~\ref{app:data_analysis} (\textbf{orange} and \textbf{blue}), or using the Power Law Integrated Curve in \cite{NANOGrav:2023ctt} (\textbf{gray}) assuming a signal-to-noise ratio (SNR) threshold of $5$ and $10$. }
\end{figure*}
\underline{Scalar field gradient} ---
It is necessary to distinguish two scenarios according to whether the retarding pressure stops the walls from accelerating before collision $\mathcal{P}_{\rm fric}
= \Delta V$ or not $\mathcal{P}_{\rm fric} \ll  \Delta V$ \cite{Ellis:2019oqb,Gouttenoire:2021kjv}.
In the later case, bubble walls run-away $\gamma_w \nearrow$, and the latent heat is dominantly kept in terms of bubble wall kinetic energy which is the main source of GWs. This occurs for very large supercooling
\begin{equation}
\label{eq:Tn_Teq_wall_acceleration}
\frac{T_{\rm n}}{T_{\rm eq}}  ~\lesssim ~ 5.3 \times 10^{-5} \left(\frac{v_\phi}{\rm 1~\rm GeV}\frac{\beta/H}{10}\frac{0.45}{g_{\rm D}} \right)^{1/4}.
\end{equation}
GWs from scalar field gradient were first computed in the so-called ``envelop'' approximation where walls are infinitely thin and collided parts are neglected \cite{Kamionkowski:1993fg,Caprini:2007xq, Huber:2008hg,Jinno:2016vai,Weir:2016tov}.
Later, collided parts were added to the computation in the so-called ``bulk flow" model at the analytical \cite{Jinno:2017fby} and numerical level \cite{Konstandin:2017sat,Lewicki:2020jiv,Lewicki:2020azd,Cutting:2020nla}. It was found that the long-lasting propagation of the infinitely thin shells produces an IR enhancement of the GW spectrum as $\Omega_{\rm PT}\propto f^{1}$ instead of $\Omega_{\rm PT}\propto f^{3}$.
For relativistic wall velocities, the bulk flow model predicts~\cite{Konstandin:2017sat}
	\begin{multline}
 \label{eq:Bulk_flow}
	 \Omega_{\rm PT}h^2 \simeq  \frac{10^{-6}}{(g_*/100)^{1/3}} \left(\frac{H_*}{\beta} \right)^{\!2} \left( \frac{\alpha}{1+\alpha} \right)^{\!2}  S_{\rm PT}(f)S_{H}(f),
	\end{multline} 
with the spectral shape $S_{\rm PT}(f)$ peaked on $f_\phi$
	\begin{equation}
	\label{eq:spectral_shape_scalar}
	S_{\rm PT}(f) = \frac{ 3(f/f_{\rm PT})^{0.9} }{2.1+0.9(f/f_{\rm PT})^{3}},\quad f_{\rm PT} = \left(\frac{a_*}{a_0}\right) 0.8 \left(\frac{\beta}{2\pi}\right),
	\end{equation}
and the redshift factor between percolation ``$*$'' and today ``$0$''
	\begin{equation}
	\label{eq:redshift_fac}
a_*/a_0 = 1.65 \times 10^{-2}~{\rm mHz}~\left(\frac{T_{\rm eq}}{100~\rm GeV}\right) \left( \frac{g_{\rm eff, \,reh}}{100} \right)^{1/6} H_{*}^{-1}.
	\end{equation}
We added the correction factor 
\begin{equation}
\label{eq:Hubble_expansion_fac}
S_{H}(f) = \frac{(f/f_{\ast})^{2.1}}{1+(f/f_{\ast})^{2.1}}, \quad f_{\ast} = c_*\left(\frac{a_*}{a_0}\right)\left(\frac{H_*}{2\pi}\right),
\end{equation}
with $c_* = \mathcal{O}(1)$ to impose an $f^{3}$ scaling for emitted frequencies smaller than the Hubble factor $ H_{\ast}/(2\pi)$ as required by causality~\cite{Durrer:2003ja,Caprini:2009fx,Cai:2019cdl,Hook:2020phx}. We fix $c_*=1$ and leave the determination of $c_*$ for future studies.

\underline{Plasma dynamics} ---
If Eq.~\eqref{eq:Tn_Teq_wall_acceleration} is not satisfied, bubble walls reach a constant Lorentz factor $\dot{\gamma}_w=0$, and the latent heat of the phase transition is dominantly transferred to the plasma, which is the main source of GWs. Friction-dominated bubble wall motion is expected to generate extremely thin and relativistic fluid configurations, which become long-lasting shock waves after bubble collisions \cite{Jinno:2019jhi}.
The large hierarchy between the bubble radius and the thickness of the shock front is a major challenge to numerical treatment. However, from a gravitational viewpoint an extremely peaked momentum distribution carried by the plasma should be indistinguishable from an extremely peaked momentum distribution carried by the scalar field. Hence we expect the GW signal in both situation to be similar. 
A second difficulty in modelling plasma dynamics is the possibility for bubble walls to be followed by relativistic shells of free-streaming particles \cite{Baldes:2020kam,Azatov:2020ufh,Gouttenoire:2021kjv,Baldes:2023fsp}, breaking down the fluid description. A recent study in the moderately relativistic regime $\gamma_w \lesssim 10$ \cite{Jinno:2022fom} suggests that the GW spectrum again resembles the one predicted in bulk flow model. 
For the two aforementioned reasons, in the present work we assume the GW signal to be given by the bulk flow model in Eq.~\eqref{eq:Bulk_flow} in the whole strongly supercooled regime $T_n \ll T_{\rm eq}$, independently of whether Eq.~\eqref{eq:Tn_Teq_wall_acceleration} is satisfied or not.\footnote{We thank Ryusuke Jinno for fruitful discussions regarding this point.}

\section{{PTA DATA ANALYSIS}}

\underline{Numerical strategy} ---

We searched for GW from 1stOPT in two open-access datasets, NG15 \cite{NANOGrav:2023gor} and IPTA2 \cite{Antoniadis:2022pcn}. The released data are presented in terms of the timing-residual cross-power spectral density $S_{ab}(f)\equiv \Gamma_{ab} h^2_c(f)/(12\pi^2)f^{-3}$, where $h_c(f)\simeq 1.26\cdot 10^{-18}(\text{Hz}/f)\sqrt{h^2\Omega_{\text{GW}}(f)}$ signifies the characteristic strain spectrum \cite{Caprini:2018mtu} and $\Gamma_{ab}$ denotes the Overlap Reduction Function (ORF) between pulsars 'a' and 'b' within a given PTA \cite{Taylor:2021yjx}.
We used the software packages ${\tt enterprise}$~\cite{enterprise} and ${\tt enterprise\_extensions}$~\cite{enterprise_ext} to compute the likelihood of observing given timing residuals assuming the presence of the SGWB from 1stOPT given in Eq.~\eqref{eq:Bulk_flow}. We used ${\tt PTMCMC}$~\cite{justin_ellis_2017_1037579} to generate the posterior distribution. For IPTA2, we marginalized over white, red and dispersion measure noises as prescribed in \cite{Antoniadis:2022pcn, Ferreira:2022zzo,Dandoy:2023jot}. For NG15, we instead used the handy wrapper ${\tt PTArcade}$ \cite{Mitridate:2023oar} with ``enterprise'' mode in which marginalization over noise parameters is automatized. We used ${\tt GetDist}$~\cite{Lewis:2019xzd} tool to plot the results.  To circumvent pulsar-intrinsic excess noise at high frequencies, the SGWB search was confined to the lowest 14 and 13 frequency bins of the NG15 and IPTA2 datasets, respectively. 
We included the BBN constraints assuming that the 1stOPT sector reheates dominantly into Standard Model degrees of freedom and, when specified, the one from PBH overproduction discussed in Sec.~\ref{sec:PBH}, to infer the prior distribution of 1stOPT parameters. Detailed information regarding data analysis and prior choices can be found in App.\ref{app:data_analysis}.

\begin{table}[h!t]
  \begin{center}
    \begin{tblr}{|Q[c,1.7cm]|Q[c,1.7cm]|Q[c,2.0cm]|Q[c,1.6cm]|}
      \hline
      \SetCell[r=2]{c}{{{\textbf{Prior} }}}&\SetCell[r=2]{c}{{{\textbf{Parameters} }}}
    & 
\SetCell[c=2]{c}{{{\textbf{Posterior mean}}}} \\
 & 
 \hline
 &
 \textrm{NG15} &
 \textrm{IPTA2}\\
      \hline \hline 
          \SetCell[r=2]{c}{{{BBN}}}& $\log_{10}{T_{\rm reh}} $ & $ -0.80^{+0.23}_{-0.23}$ & $-1.34^{+0,3}_{-0.3}$   \\\hline 
                   & $\beta/H$& $ 9.8^{+4.0}_{-2.5}$ &$13.8^{+6.6}_{-4.2}$   \\
       \hline \hline  
          \SetCell[r=2]{c}{{{BBN + PBH}}}& $\log_{10}{T_{\rm reh}} $ & $ -0.86^{+0.37}_{-0.26}$ & $-1.36^{+0.38}_{-0.32} $   \\\hline 
                   & $\beta/H$& $ 10.7^{+2.8}_{-3.3}$ &$14.1^{+5.8}_{-3.6}$   \\
     \hline \hline 
          \SetCell[r=3]{c}{{{BBN + PBH + SMBH }}}& $\log_{10}{T_{\rm reh}} $ & $ -0.92^{+0.70}_{-0.39}$ & $ -1.4^{+1.3}_{-1.3} $   \\\hline 
                   & $\beta/H$& $ 12.6^{+8.3}_{-6.1}$ &$29^{+161}_{-23}$   \\\hline 
                   & $\log_{10}{A_{\rm SMBH}}$& $ -15.5^{+1.1}_{-0.78} $ &$-14.70^{+0.36}_{-0.12}$   \\
         \hline \hline 
          SMBH alone& $\log_{10}{A_{\rm SMBH}} $ & $ -14.62^{+0.11}_{-0.12}$ & $-14.46^{+0.07}_{-0.05}$    \\
      \hline
    \end{tblr}
    \caption{\label{tab:meanPoseterior} Mean parameter values with 68$\%$ confidence interval of the probability distribution for distinct prior information (\textbf{rows}) and observed data (\textbf{columns}). The GW spectrum from 1stOPT is taken from Bulk Flow model Eq.~\eqref{eq:Bulk_flow} in the supercooled limit $\alpha \gg 1$. }
  \end{center}
\end{table}

\underline{Supercooled PT} ---
We conducted searches for GW from strong 1stOPT ($\alpha \gg 1$) in isolation, GW from SMBH binaries individually, as well as a combined analysis of 1stOPT and SMBH binaries. In Fig.~\ref{fig:1stOPT_trapeze}, we show the GW spectra with parameters set to their mean posterior values given in Tab.~\ref{tab:meanPoseterior}.
The $68\%$ and $95\%$ confidence contours are depicted in Fig.~\ref{fig:betaOH_TnOTeq}-left. The posterior for the combined analysis of 1stOPT and SMBH is reported to Figs.~\ref{fig:1stOPT_trapeze_join} and \ref{fig:beta_vs_Treh_PTA_SMBH} in the appendix. We assumed a flat prior on the strain amplitude of the SGWB from SMBH binaries, as well as the spectral slope of $13/3$ associated with GW-driven inspirals.
To quantify the evidence provided by the observed PTA data, denoted as $\mathcal{D}$, in favor of one model, say $X$, versus another, say $Y$, we employ the Bayesian factor
\begin{equation}
    \textrm{BF}_{Y,X} \equiv \mathcal{P}(\mathcal{D}|Y)\,/\,\mathcal{P}(\mathcal{D}/X),
\end{equation}
which we compute using the product-space sampling method \cite{Taylor:2021yjx}
implemented in ${\tt enterprise\_extensions}$ \cite{enterprise_ext}. Here, $\mathcal{P}(\mathcal{D}/X)$ is the likelihood probability of observing data D given the model X.  The outcomes of the Bayesian model comparison presented in Tab.~\ref{tab:BF}, according to Jeffrey's scale \cite{1939thpr.book.....J,kass1995bayes}, suggests that NG15 data `substantially' favours the presence of a GW signal from 1stOPT aside to the one from SMBHB. Instead, IPTA2 data remains inconclusive.

\begin{table}[h!t]
  \begin{center}
    \begin{tblr}{|Q[c,1.cm]|Q[c,2.cm]|Q[c,1.3cm]|Q[c,0.8cm]|Q[c,0.8cm]|}
      \hline
      \SetCell[r=2]{c}{{{\textbf{Model X} }}}
    & \SetCell[r=2]{c}{{{\textbf{Model Y} }}}& \SetCell[r=2]{c}{{{\textbf{Prior} }}}& 
\SetCell[c=2]{c}{{{${\textrm{BF}_{Y,X}}$}}} \\
 & &&
 \hline
 \textrm{NG15} &
 \textrm{IPTA2}\\
        \hline
        \SetCell[r=2]{c=1}{{{SMBH}}} & \SetCell[r=2]{c=1}{{{1stOPT}}}&BBN & 24 & 0.50 \\
        \hline
        & &BBN + PBH & 15 & 0.49 \\
        \hline
         SMBH & SMBH+1stOPT & BBN+PBH & 9.3 & 1.2 \\
      \hline
    \end{tblr}
    \caption{\label{tab:BF} Bayesian factors $\rm BF_{Y,X}$ with values significantly exceeding 1 indicate support for interpretation $Y$ with respect to $X$. Conversely, values approaching 1 suggests no discernible preference between $X$ and $Y$. We can see that the 1stOPT interpretation is favored with respect to SMBH binaries in NG15 data and that the PBH prior only slightly worsens the fit.}
  \end{center}
\end{table}

\underline{Exclusion bounds} ---
Under the assumption that the PTA signal does not arise from 1stOPT, we have derived upper limits on the GW signal emanating from 1stOPT. As depicted in Fig.~\ref{fig:betaOH_TnOTeq}-right, these limits correspond to lower bounds on the rate of completion, going up to $\beta/H \lesssim 20$.
As discussed in App.~\ref{app:data_analysis}, these lower limits are conservative as the GW spectrum from SMBH was not included in the analysis.

\begin{figure*}[ht!]
    \centering
    \hspace{-0.7cm}
    \includegraphics[width=0.7\textwidth]{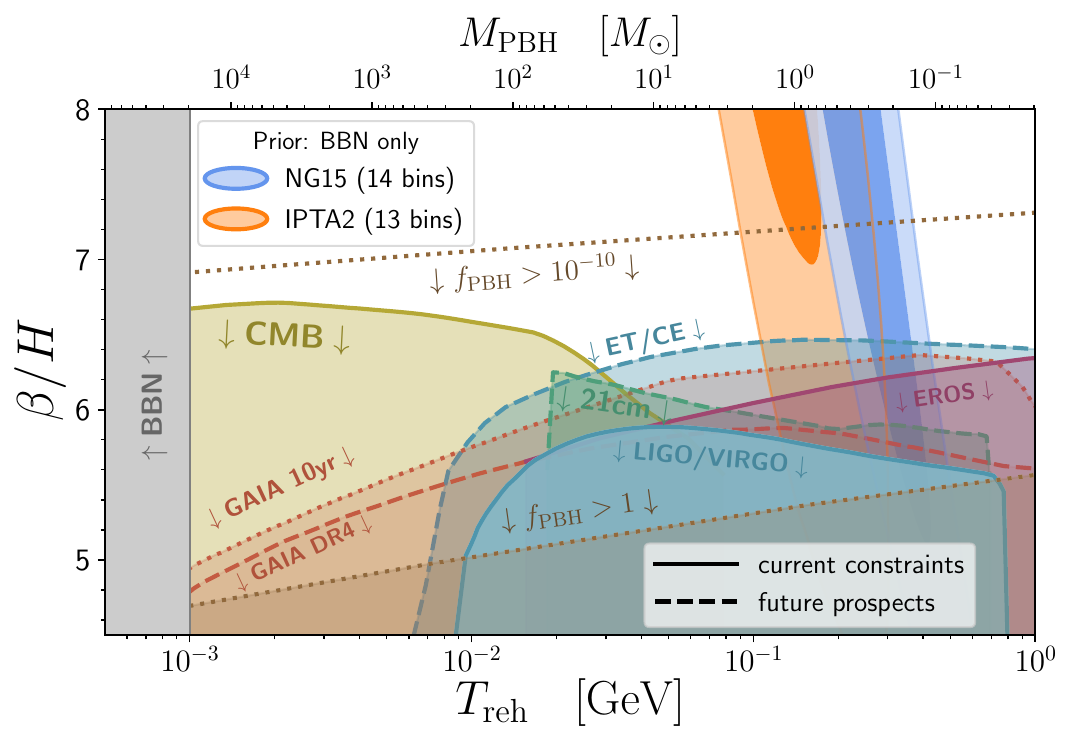}
    \caption{\label{fig:beta_vs_Treh_allconstraints} \small  The \textbf{ellipses} are the posterior distributions obtained after a Bayesian search of SGWB sourced by a supercooled 1stOPT in NG15 and IPTA2 data sets. We overlay the region producing PBHs detectable by different observatories, see Sec.~\ref{sec:PBH} for details.}
\end{figure*}

\section{PRIMORDIAL BLACK HOLES}
\label{sec:PBH}
\underline{Supercooled late-blooming mechanism} ---

In \cite{Gouttenoire:2023naa}, it was demonstrated that PBHs could be produced in observable amount during supercooled PT through a process termed ``late-blooming''.
During 1stOPT, the nucleation sites of bubbles are randomly dispersed across the entire volume of the false vacuum. As the universe gets close to the point of percolation, there remains a non-zero probability of identifying Hubble-sized regions where nucleation has not yet initiated.
Throughout the supercooled PT, these delayed regions maintain a constant vacuum energy, while the energy density in their vicinity redshifts like radiation. Upon completion of percolation, these ``late-bloomers'' evolve into over-dense regions. If these regions are Hubble-sized and exceed a certain density threshold $\delta \rho/\rho \gtrsim 0.45$, they collapse into PBHs.
We direct the reader to \cite{Gouttenoire:2023naa} for the precise analytical formula to estimate the abundance and mass of those PBHs.\footnote{Some other works \cite{Kodama:1982sf,Lewicki:2023ioy,Liu:2021svg,Kawana:2022olo} find a different PBH abundance. Refs.~\cite{Kodama:1982sf,Lewicki:2023ioy,Kawana:2022olo} find a lower PBH abundance because the formalism is restricting collapsing patch to remain $100\%$ vacuum dominated until collapse. Ref.~\cite{Liu:2021svg} find a larger abundance because  nucleation is not accounted in the entire past light-cone of a collapsing patch. Instead, Ref.~\cite{Gouttenoire:2023naa} accounts for nucleation to take place not only in the whole past light-cone but also in the collapsing patch itself as long as the critical overdensity is reached. 
Finally, another study Ref.~\cite{Baldes:2023rqv} confirms, with an appreciable level of detail, the findings of  \cite{Gouttenoire:2023naa} used in the present work.
}
The mass distribution of those PBHs, left for future studies in \cite{Gouttenoire:2023naa}, is assumed to resemble a delta function in the present work.
We included the PBH overproduction constraints as a prior in the Bayesiasn analysis. The Bayes factors shown in Tab.~\ref{tab:BF} is unaffected for IPTA2 and only decreases from $24$ to $15$ for NG15.
We have plotted the contour lines representing the PBH fraction of dark matter $f_{\rm PBH}$ in Fig.~\ref{fig:betaOH_TnOTeq} and the PBH mass in Fig.~\ref{fig:beta_vs_Treh_allconstraints}. In addition, we overlay cosmological and astrophysical constraints on this population of PBHs.

\underline{Excluded regions and detection prospects} ---
With solid lines, we show current constraints.
In yellow, we have the exclusion regions arising from distortion of the Cosmic Microwave Background (CMB) caused by X-rays from accretion which modify the ionization history between recombination and reionizaton \cite{Ali-Haimoud:2016mbv,Poulin:2017bwe,Serpico:2020ehh}.
In purple, we show the constraints using the search for photometric magnification (strong lensing) of stars in the Magellanic clouds conducted on Eros data~\cite{EROS-2:2006ryy}.
The solid cyan-colored region represents constraints derived from the data collected by LIGO/Virgo interferometers \cite{Nakamura:1997sm,Raidal:2018bbj,Kavanagh:2018ggo,LIGOScientific:2019kan,DeLuca:2020qqa}.
With dashed lines, we show future prospects. In green, we have the reach of 21 cm surveys due to heating and ionization of the intergalactic medium via X-rays produced during accretion \cite{Mena:2019nhm,Villanueva-Domingo:2021cgh,Villanueva-Domingo:2021spv}. In red, we have the forecast from the search for transient astrometric deviation (weak lensing) of single or multiple stars in GAIA time-series data \cite{Chen:2023xyj,VanTilburg:2018ykj,Verma:2022pym}. Finally, in dashed cyan we show the prospect for detecting GW from PBH binaries with Einstein telescope and Cosmic Explorer \cite{Chen:2019irf,Pujolas:2021yaw}.

\section{CONCLUSION}

We conducted a Bayesian analysis of the NANOGrav 15-yr (NG15) and IPTA DR2 (IPTA2) timing residuals. Our findings indicate that NG15 indicate a substantial preference for the presence of a strong first-order phase transitions (1stOPT) in isolation or combined with SGWB from SMBH binaries, while IPTA2 remains inconclusive on which scenario is preferred. 
The phase transition is characterized by a remarkably low completion rate, e.g. $\beta/H\simeq 12.6$ and $10.7$ for NG15 with and without astrophysical signal from SMBH binaries. From a theoretical perspective, such a value is typical of supercooled phase transitions, characterized by a strong first-order phase transition with a parameter $\alpha$ significantly larger than 1, e.g. \cite{Caprini:2015zlo,Caprini:2019egz,Gouttenoire:2022gwi}, which motivates the choice of prior $\alpha \gg 1$ done in this work.  These cosmological scenarios have been demonstrated to produce primordial black holes (PBHs) in considerable quantities when $\beta/H \lesssim 7$ \cite{Gouttenoire:2023naa}.
The Bayes factor of the strong 1stOPT interpretation with respect to SMBH binary one is only reduced from $24$ to $15$ in NG15 after including the PBH prior, while it is not affected in IPTA2.
 
 However, we showed that the 1stOPT interpretation of the PTA signal might be associated with the presence of solar-mass PBHs in our universe today. We further assessed the potential for detecting these PBHs using different observational techniques, including 21 cm cosmological hydrogen line observations, astrometry with the GAIA mission and next-generation kilohertz frequency GW interferometers such as the Einstein Telescope (ET) and Cosmic Explorer (CE).
 We conclude that 1stOPTs can be ranged alongside domain wall networks \cite{Gouttenoire:2023ftk,Gouttenoire:2023gbn} and scalar induced GW \cite{Chen:2019xse,Dandoy:2023jot,Franciolini:2023pbf} in the category of the early-universe interpretations of PTA signal capable of producing multi-solar-mass PBHs in quantities that are potentially observable.
 
In the event that an astrophysical explanation becomes definitive, we established 68$\%$ and 95$\%$ exclusion constraints on the parameter space of 1stOPT, up until $\beta/H\gtrsim 20$. Under these conditions, it would effectively preclude any possibility of detecting PBHs from supercooled PTs within the mass range $[1~M_{\odot},~10^3~M_{\odot}]$. 

We must emphasize that our current comprehension of the
GW spectrum resulting from supercooled phase transitions is
still in its early stages. The assumptions are founded on the
bulk flow model, in which GWs are sourced by the expansion of an infinitely thin distribution of the stress-energy momentum tensor. Future investigations are necessitated to probe
potential modifications of the GW spectrum that could be induced by non-linear effects, such as those arising from rel-
ativistic shock waves, or deviations from a fluid description.

Finally, we would like to note that besides GW and PBH signatures, supercooled phase transitions can be efficient in producing ultra-relativistic particles around bubble walls \cite{Baldes:2020kam,Azatov:2021ifm,Gouttenoire:2021kjv,Jinno:2022fom} which can source dark matter \cite{Baldes:2021aph,Azatov:2021ifm,Baldes:2022oev,Baldes:2023fsp} or baryonic asymmetry \cite{Azatov:2021irb,Baldes:2021vyz}.

{\bf Acknowledgements.}---%
The author is grateful to Iason Baldes, Ryusuke Jinno, Marius Kongsore, Fabrizio Rompineve, Miguel Vanvlasselaer and Tomer Volansky for fruitful discussions and to the Azrieli Foundation for the award of an Azrieli Fellowship. This work was conducted using the high performance computing cluster resources of Tel Aviv University.

\appendix

\clearpage
\appendix
\onecolumngrid

\fontsize{11}{13}\selectfont

\titleformat{\section}
{\normalfont\fontsize{12}{14}\bfseries  \centering }{\thesection.}{1em}{}
\titleformat{\subsection}
{\normalfont\fontsize{12}{14}\bfseries \centering}{\thesubsection.}{1em}{}
\titleformat{\subsubsection}
{\normalfont\fontsize{12}{14}\bfseries \centering}{\thesubsubsection)}{1em}{}

\titleformat{\paragraph}
{\normalfont\fontsize{12}{14}\bfseries  }{\thesection:}{1em}{}

\begin{figure*}[ht!]
    \centering
    \includegraphics[width=0.45\textwidth]{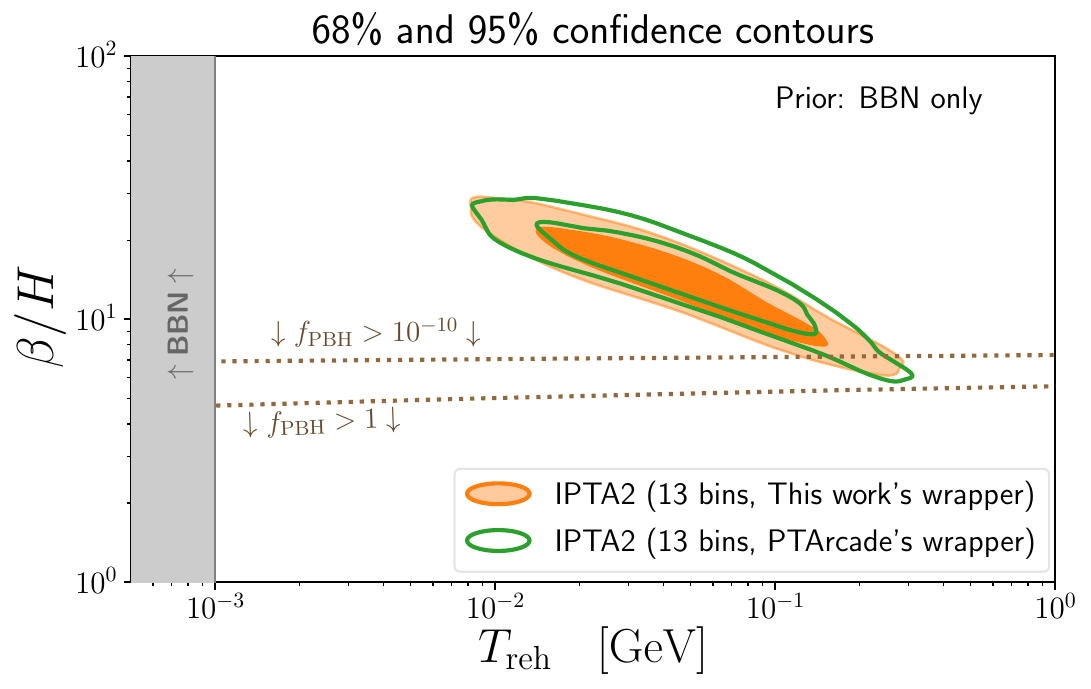}
    \includegraphics[width=0.45\textwidth]{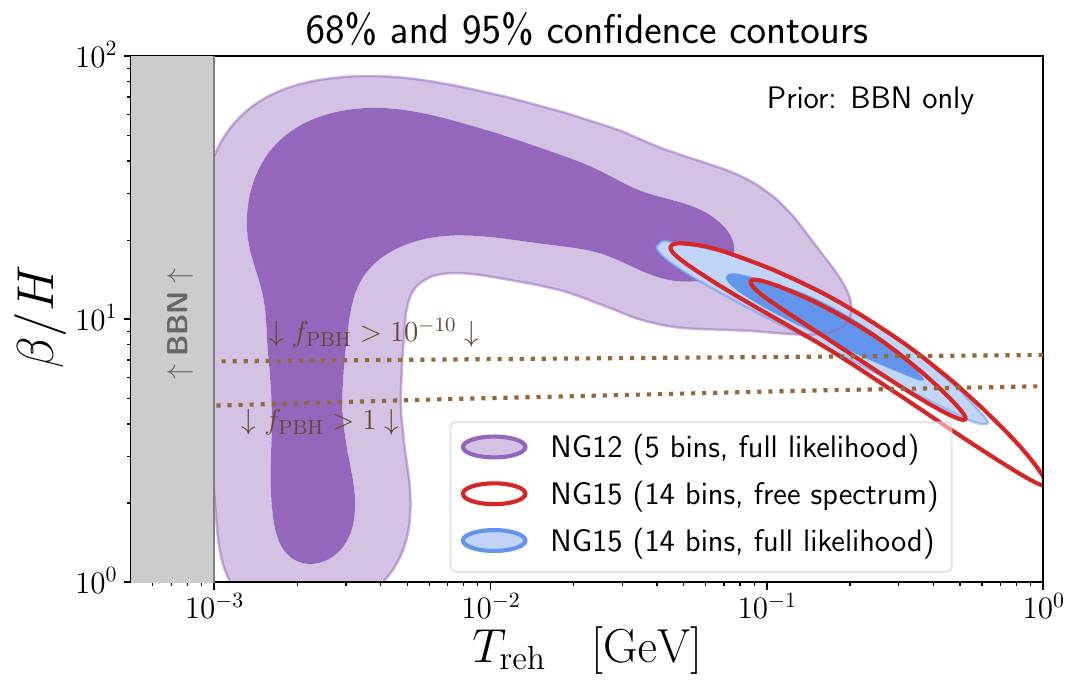}
      \caption{\label{fig:ceffyl_vs_free_spe} \small  \small    \textbf{Left}: We compare the posterior resulting from the Bayesian analysis realized with our own wrapper of ${\tt enterprise\_extensions}$ \cite{enterprise_ext} as described in App.~\ref{app:data_analysis} (\textbf{orange}) with the one using the public software ${\tt  PTArcade}$ \cite{Mitridate:2023oar} (\textbf{green}) . \textbf{Right}: Numerous analysis in the literature were performed with NANOGrav 12.5-yr (NG12) with 5 frequency bins (\textbf{purple}), which is very different from the recent NG15 14-bin posterior (\textbf{blue}). We also compare the posterior obtained with the fast method of \cite{Lamb:2023jls}, called ``ceffyl'', which relies on a fit of the violin plot (\textbf{red}). All posteriors in the right panel were generated with ${\tt  PTArcade}$ \cite{Mitridate:2023oar}.    }
\end{figure*}

\section{Data analysis}
\label{app:data_analysis}

The purpose of this Appendix is to delineate the Bayesian search methodology employed in our study. We started rely on the NG15 dataset~\cite{NG15yrdata} and on Version B of the IPTA2 dataset~\cite{IPTADR2data}.  To ascertain noise parameters of IPTA2, we closely follow the approach adopted by IPTA collaboration \cite{Antoniadis:2022pcn}, see also \cite{Ferreira:2022zzo,Dandoy:2023jot}. We then checked that we obtained consistent result with the software ${\tt PTArcade}$ \cite{Mitridate:2023oar} in which noise marginalization has been automatised, see Fig.~\ref{fig:ceffyl_vs_free_spe}-left. Instead the Bayesian analysis of NG15 was done solely with the ``enterprise'' mode of  ${\tt PTArcade}$ \cite{Mitridate:2023oar}.  We perform the search for SGWB in the first 13 and 14 frequency bins of IPTA2 and NG15, respectively. 

\underline{IPTA2 analysis.} ---
We now describe the Bayesian analysis of IPTA2 which we performed ourselves without the use of ${\tt PTArcade}$ \cite{Mitridate:2023oar}.
We adapted the software packages ${\tt enterprise}$~\cite{enterprise} and ${\tt enterprise\_extensions}$~\cite{enterprise_ext} to incorporate GW spectra from 1stOPT in terms of the power spectrum in timing residual, and used them to compute the likelihood function, symbolized as $\mathcal{P}(\mathcal{D}|\theta)$. This function encapsulates the probability of observing the data $\mathcal{D}$ given a specific set of model parameters $\theta$. The posterior distribution, $\mathcal{P}(\theta|\mathcal{D})$, which illustrates the probability distribution of model parameters $\theta$ given the observed data $\mathcal{D}$, is linked to the likelihood function via Bayes's theorem
 \begin{equation}
     \mathcal{P}(\theta|\mathcal{D}) = \frac{\mathcal{P}(\mathcal{D}|\theta)\mathcal{P}(\theta)}{\mathcal{P}(\mathcal{D})}.
 \end{equation}
 Within this equation, $\mathcal{P}(\theta)$ is the prior distribution, representing preliminary knowledge of the parameters prior to data observation, while $\mathcal{P}(\mathcal{D})$ is the marginal likelihood or evidence, functioning as a normalization constant to ensure that the posterior distribution integrates to 1.
The parallel-tempering Markov Chain Monte-Carlo sampler ${\tt PTMCMC}$~\cite{justin_ellis_2017_1037579} was employed to reconstruct the posterior distribution $\mathcal{P}(\theta|\mathcal{D})$ using an enhanced version of the Metropolis-Hastings algorithm \cite{Taylor:2021yjx}. The ${\tt GetDist}$ tool~\cite{Lewis:2019xzd} was subsequently used to plot the posterior distributions and upper limits.
The pulsar noise parameters employed in the likelihood function can be classified into three distinct categories: white noise, red noise, and dispersion measures (DM). The white noise parameters are grouped into three sets for each backend/receiver associated with a given pulsar: EFAC ($E_k$), EQUAD ($Q_k[s]$), and ECORR ($J_k[s]$). The values of the white noise parameters are fixed to the mean posterior values obtained by performing single pulsar analysis devoid of GW signals. We only kept pulsars with more than 3 years of observation time which corresponds to 53 pulsars. Instead the Bayesian analysis of NG15 data performed via ${\tt PTArcade}$ contains $68$ pulsars with more than $3$ years of observation. We employ the Jet Propulsion Laboratory Development Ephemeris DE438 and the Terrestrial Time  reference timescale of the International Bureau of Weights and Measures BIPM18.
Next, for the multi-pulsar analysis incorporating the GW signals, we account for two power-law red noise parameters per pulsar, specifically the amplitude at the reference frequency of $\text{yr}^{-1}$ denoted as $A_{\text{red}}$, and the spectral index denoted as $\gamma_{\text{red}}$. Additionally,  we incorporate power-law errors associated with dispersion measures (DM). We note that the treatment of DM noise as a Gaussian process is specific of IPTADR2 dataset. Instead, in the analysis of NG15 data performed via ${\tt PTArcade}$, but also in the analysis of NANOGrav 12.5-year (NG12) done in \cite{NANOGrav:2020spf}, pulse dispersion is modelled by a set of ``per- epoch'' parameters describing the DM offset from a nominal fixed value \cite{NANOGrav:2015qfw,Jones:2016fkk}. These can add dozens of additional parameters per pulsar \cite{NANOGrav:2023hde}.
In the individual pulsar analysis of PSR J1713+0747 (in IPTA2 but also in NG15), we extend our consideration to encompass a DM exponential dip parameter, following the methodology described in \cite{Antoniadis:2022pcn}. The priors for the noise parameters are reported in Tab.~\ref{tab:priors}, along with the priors for the parameters for the GW spectra from 1stOPT and SMBH binaries. 
To economize on computational time, we adopt the methodology of previous studies \cite{NANOGrav:2020bcs,Antoniadis:2022pcn} and in our search for a GW background we utilize only auto-correlation terms $I=J$ in the Overlap Reduction Function (ORF) $\Gamma_{IJ}$, rather than the complete Hellings-Downs ORF with $I\neq J$.
 We acquire $10^6$ samples per analysis presented in this study and discard 25$\%$ of each chain as burn-in. We could replicate the posteriors of \cite{NANOGrav:2020bcs} and \cite{Antoniadis:2022pcn} for a power-law model with excellent concurrence. 

\begin{figure*}[ht!]
    \centering
    \includegraphics[width=0.7\textwidth]{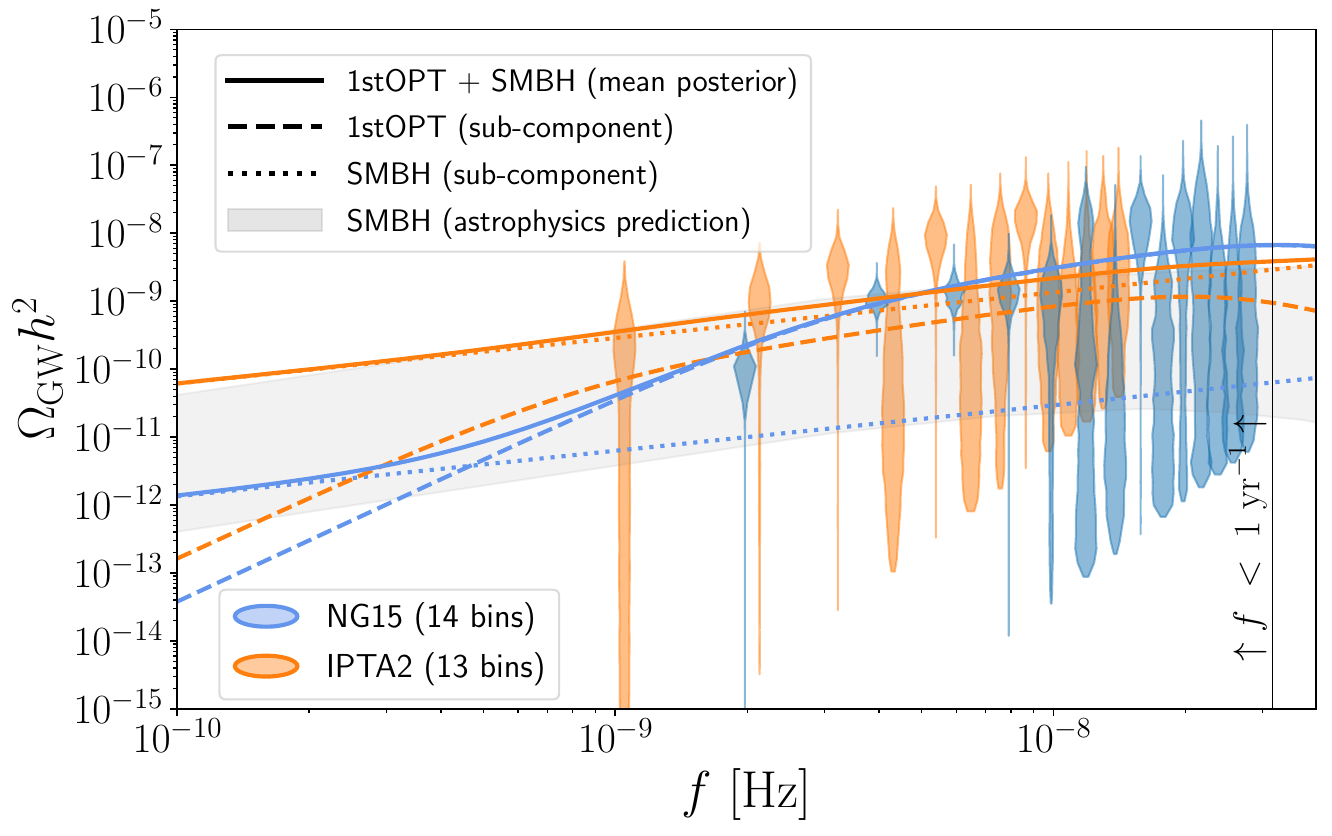}
      \caption{\label{fig:1stOPT_trapeze_join} \small  \small    We show the combined GW signal from 1stOPT and SMBH binaries with mean posterior values for the rate of completion $\beta/H$ and reheating temperature $T_{\rm reh}$ (\textbf{blue} and \textbf{orange}). The \textbf{gray} band shows the SGWB from the incoherent superposition of a population of Monte-Carlo-simulated SMBH binaries \cite{Rosado:2015epa}. The band brackets $90\%$ of the simulated population. }
\end{figure*}

  The violin features shown in Figs.~\ref{fig:1stOPT_trapeze} and \ref{fig:1stOPT_trapeze_join} are obtained with the free-spectrum approach described in \cite{Chalumeau:2021fpz}. We do not repeat this analysis and instead take the data directly from \href{https://zenodo.org/record/8060824}{NG15} and
\href{https://zenodo.org/record/5787557}{IPTA2}.

Our study encompasses two types of analyses. The first, a detection analysis, identifies the region of parameter space in which GWs from 1stOPT can account for the common-spectrum process in the datasets. Here, we use a uniform prior on the logarithm of each parameter and adopt a prior on $\beta/H$ due to the BBN bound and - when mentioned - PBH overproduction. 
The second, an lower-limit analysis, seeks to constrain the rate of completion of the phase transition $\beta/H$.
There, we use a uniform prior on $H/\beta$ instead of $\log_{10}(\beta/H)$ as described in \cite{Romano:2016dpx,Taylor:2021yjx}. We made the conservative choice to not include the GW spectrum from SMBH in the lower-limit analysis, see the related discussion in \cite{Dandoy:2023jot}. All prior choices are given in Tab.~\ref{tab:priors}.

\begin{table}[h!t]
  \begin{center}
    \begin{tblr}{|Q[c,2.cm]|Q[c,1.5cm]|Q[c,3cm]|Q[c,2.5cm]|Q[c,2.5cm]|Q[c,2.5cm]|}
   \hline  \bf{Signal} & \bf{Parameter}  &{\bf{Description}} & \SetCell[c=2]{c}{{\bf{Prior}}}& & {\bf{Comments}} \\
      \hline \hline
\SetCell[r=3]{c}{{{White Noise}}}& $E_{k}$ & EFAC per backend/receiver system & \SetCell[c=2]{c}{{Uniform $[0, 10]$ }}&& single-pulsar analysis only \\\hline
& $Q_{k}$ [s] & EQUAD per backend/receiver system & \SetCell[c=2]{c}{{log-Uniform $[-8.5, -5]$}} & & single-pulsar analysis only \\\hline
& $J_{k}$ [s] & ECORR per backend/receiver system & \SetCell[c=2]{c}{{log-Uniform $[-8.5, -5]$}} & & single-pulsar analysis only  \\\hline \hline
\SetCell[r=2]{c}{{{Red Noise}}} &$A_{\rm red}$ & red-noise power-law amplitude & \SetCell[c=2]{c}{{log-Uniform $[-20, -11]$}}& & one parameter per pulsar  \\\hline
& $\gamma_{\rm red}$ & red-noise power-law spectral index & \SetCell[c=2]{c}{{log-Uniform $[0, 7]$}}& & one parameter per pulsar \\
\hline\hline
\SetCell[r=2]{c}{{{DM Noise}}}& $A_{\rm DM}$ & DM noise power-law amplitude & \SetCell[c=2]{c}{{log-Uniform $[-20, -11]$}}& & one parameter per pulsar (IPTA DR2)  \\\hline
& $\gamma_{\rm DM}$ & DM noise power-law spectral index & \SetCell[c=2]{c}{{log-Uniform $[0, 7]$}}& & one parameter per pulsar (IPTA DR2) \\
\hline\hline
\SetCell[r=3]{c}{{{1stOPT (confidence contours)}}}& $T_{\rm reh}~\rm [GeV]$ & PT temperature & \SetCell[r=3]{c}{{{ When specified: \\ 1) BBN prior in Eq.~\eqref{eq:Delta_Neff} \\2) PBH prior in Eq.~\eqref{eq:PBH_prior}}}}& log-Uniform $[-4, 1]$  & one parameter per PTA dataset \\\hline
&$\beta/H$ & bubble nucleation rate & &log-Uniform $[0, 3]$& one parameter per PTA data et \\\hline
&$\alpha$ & PT strength && log-Uniform $[-2, 2]$ or $\alpha = +\infty$ & one parameter per PTA dataset \vspace{-0.5cm} \\\hline
\hline
\SetCell[r=3]{c}{{{1stOPT (exclusion contours)}}}& $T_{\rm reh}~ \rm [GeV]$ & PT temperature &  \SetCell[c=2]{c}{{{fixed (analysis run over a grid in  $[10^{-3},10^1]$)}}}&  & one parameter per PTA dataset \\\hline
&$\beta/H$ & bubble nucleation rate &  \SetCell[c=2]{c}{{Uniform $[10^{-2}, 1]$ on $H/\beta$}}&& one parameter per PTA dataset \\\hline
&$\alpha$ & PT strength &  \SetCell[c=2]{c}{{$\alpha = +\infty$}} & one parameter per PTA dataset \\\hline
\hline
\SetCell[r=2]{c}{{{SMBH}}}&$A_{\mathrm{SMBH}}$ & SMBH strain amplitude & \SetCell[c=2]{c}{{log-Uniform $[-17.5, -12.5]$}} & & one parameter per PTA dataset \\\hline
&$\gamma_{\mathrm{SMBH}}$ & SMBH power-law spectral index &\SetCell[c=2]{c}{{$\gamma_\mathrm{SMBH}=13/3$}}& & fixed \\\hline
    \end{tblr}
    \caption{\label{tab:priors} Prior assumptions on the parameters used in the Bayesian analysis of this work. }
  \end{center}
\end{table}

\underline{BBN prior.} ---
 As a sub-component of the total energy density of the universe, the latent heat $\Delta V$ can impact the expansion rate of the universe which is strongly constrained by BBN and CMB. Its effect can be encoded in the effective number of extra neutrino relics
\begin{equation}
\label{eq:Delta_Neff}
 N_{\rm eff} = \frac{8}{7}\left( \frac{\rho_{\rm tot}-\rho_{\gamma}}{\rho_\gamma} \right)\left( \frac{11}{4} \right)^{4/3},
\end{equation}
where $\rho_{\gamma}$ is the photon number density. The total number of effective degrees is constrained by CMB measurements \cite{ParticleDataGroup:2022pth} to $N_{\mathsmaller{\rm eff}} = 2.99_{-0.33}^{+0.34}$ and by BBN predictions \cite{Mangano:2011ar, Peimbert:2016bdg} to $N_{\mathsmaller{\rm eff}} = 2.90_{-0.22}^{+0.22}$ whereas the SM prediction \cite{Mangano:2005cc, deSalas:2016ztq} is $N_{\mathsmaller{\rm eff}}  \simeq 3.045$. The latent heat parameter of a generic 1stOPT reads
\begin{equation}
\label{eq:alpha_tot_app}
    \alpha = \frac{\rho_{\rm DW}(T)}{\frac{\pi^2}{30}g_*(T)T^4},
\end{equation}
where $T$ is the photon temperature and $g_*(T)$ contains eventual dark degrees of freedom. The maximal contribution to $N_{\rm eff}$ occurs at reheating after percolation
\begin{equation}
\label{eq:Delta_Neff_bound}
    \Delta N_{\rm eff}(T) =\frac{8}{7}\left( \frac{{g_{*}(T)}}{2} \right)\left( \frac{11}{4} \right)^{4/3}\alpha(T).
\end{equation}
The BBN bound $\Delta N_{\rm eff}  \lesssim 0.3$ \cite{Pitrou:2018cgg,Dvorkin:2022jyg} applies after neutrino decouples below the temperature $T_{\rm dec}$ where $g_*(T< T_{\rm dec})\equiv 2+(7/8)\cdot 6\cdot (4/11)^{4/3} \simeq 3.36$. We obtain
\begin{equation}
\label{eq:Delta_N_eff_app}
    \Delta N_{\rm eff} = 7.4~\alpha~\lesssim~0.3,
\end{equation}
Two scenarios must be distinguished. The first one is when reheating after percolation occurs in a dark sector, in which case Eq.~\ref{eq:Delta_N_eff_app} is the BBN constraints. The second one is when reheating after the 1stOPT occurs into the Standard Model, in which case Eq.~\ref{eq:Delta_N_eff_app} applies only if the reheating temperature is below the neutrino decoupling temperature $T_{\rm reh} \lesssim 1~\rm MeV$. The last case is the scenario we consider in this work.
Note that stronger BBN constraints have been considered in the literature, e.g. $T_{\rm reh} \lesssim 3~\rm MeV$ in \cite{Bai:2021ibt}, or $T_{\rm reh} \lesssim 2~\rm MeV$ and  $4~\rm MeV$ for cases of electromagnetic and hadronic decays respectively \cite{Kawasaki:2000en,Hasegawa:2019jsa}.

\underline{PBH prior.} ---
The condition of not producing PBH with an energy density larger than the one of observed dark matter, $f_{\rm PBH} < 1$, implies a lower bound on the rate of completion of a 1stOPT \cite{Gouttenoire:2023naa}
\begin{equation}
\label{eq:PBH_prior}
\beta/H ~\gtrsim~ \left(5.54+0.232 \log_{10}\left(\frac{T_{\rm reh}}{\rm GeV}\right)-0.00512 \log_{10}^2\left(\frac{T_{\rm reh}}{\rm GeV}\right)\right)\left( 1 - 0.0695\ln\left( 1+\frac{908.1}{\alpha^{3.204}}\right) \right),
\end{equation}
where we have introduced an analytical function fitted on numerical results of \cite{Gouttenoire:2023naa}.  
When specified, we include the constraint in Eq.~\eqref{eq:PBH_prior} as prior information on $\beta/H$ and $T_{\rm eq}$. Due to the exponential dependence of the PBH abundance on $\beta/H$, the precise PBH constraints due to astrophysical and cosmological constraints, as shown in e.g. Fig.~\ref{fig:beta_vs_Treh_allconstraints}, make little difference with respect to simple criterion $f_{\rm \mathsmaller{PBH}} <1$.

\begin{figure*}[ht!]
    \centering
    \includegraphics[width=0.9\textwidth]{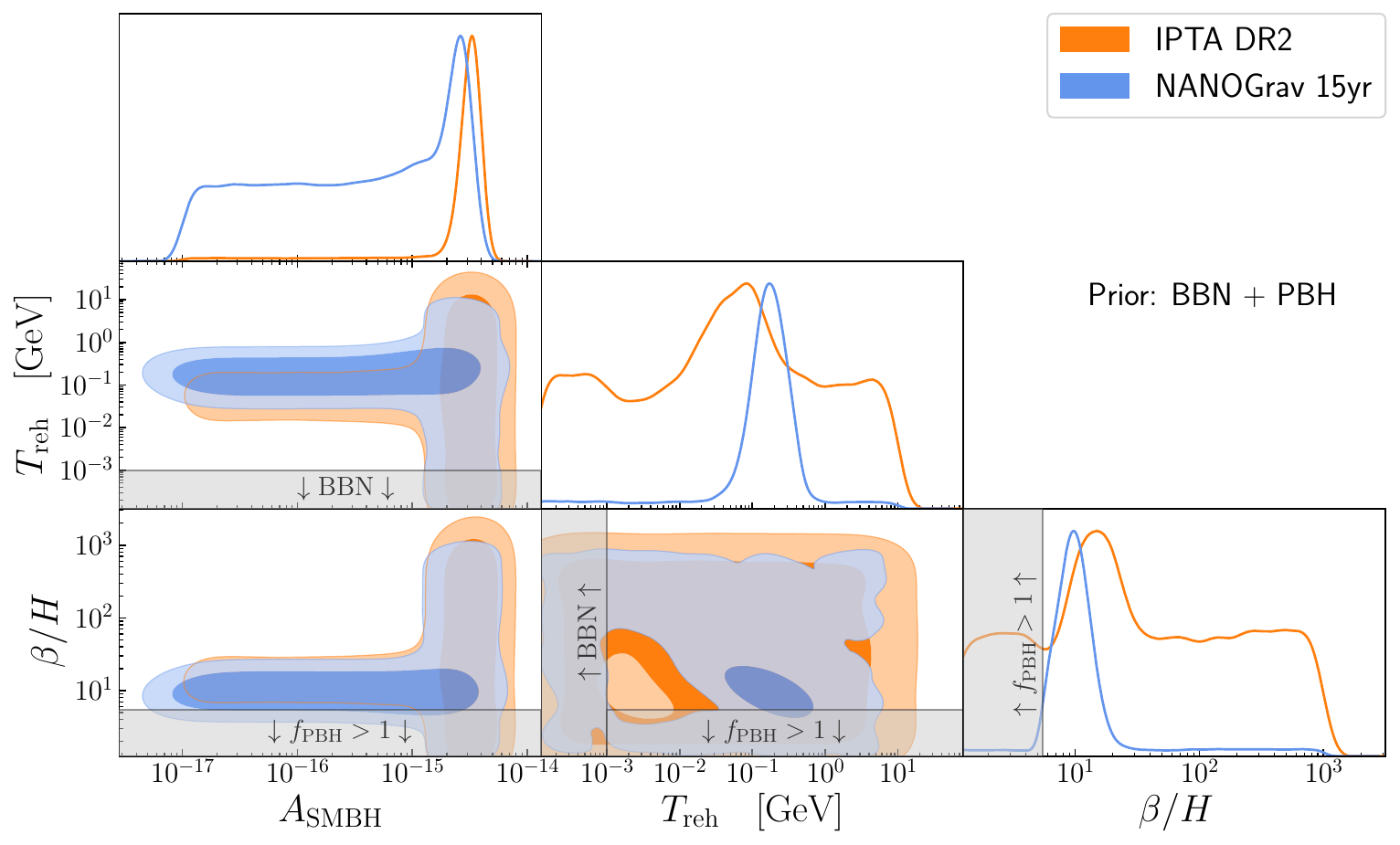}
    \caption{\label{fig:beta_vs_Treh_PTA_SMBH} We performed a Bayesian analysis of a superposition of SGWB from 1stOPT and SMBH binary mergers.}
\end{figure*}

\section{Combined GW from 1stOPT and SMBH binaries}

The characteristic strain spectrum of a population of circular GW-driven SMBH binaries is a red-tilted power-law \cite{Phinney:2001di}
\begin{equation}
    h_c(f) = A_{\rm SMBH} \left(\frac{f}{1~\rm yr^{-1}}\right)^{-2/3},
\end{equation}
where $A_{\rm SMBH}$ is the strain amplitude at $ 1~\rm yr^{-1} \simeq 3.2\times 10^{-8}$. In terms of the fractional energy density, it corresponds to the blue-tilted power-law
\begin{equation}
\Omega_{\rm SMBH}(f) = \frac{2\pi^2}{3H_0^2}f^2 h_c^2(f) \propto f^{2/3}.
\end{equation}
We conduct search for combined GW from both supercooled 1stOPT and SMBH binaries. We present the posterior distribution of model parameters $(A_{\rm SMBH}, T_{\rm reh}, \beta/H)$ in Fig.~\ref{fig:beta_vs_Treh_PTA_SMBH}. We included BBN and PBH constraints in the prior distribution of 1stOPT parameters. The mean posterior values of the parameters are reported in Tab.~\ref{tab:meanPoseterior} and the associated GW spectra are plotted in Fig.~\ref{fig:1stOPT_trapeze_join}.

\bibliography{biblio}






\renewcommand{\tocname}{\Large  Table of contents
\vspace{1 cm}}%

\titleformat{\section}
{\normalfont\fontsize{12}{14}\bfseries  \centering }{\thesection.}{1em}{}
\titleformat{\subsection}
{\normalfont\fontsize{12}{14}\bfseries \centering}{\thesubsection.}{1em}{}
\titleformat{\subsubsection}
{\normalfont\fontsize{12}{14}\bfseries \centering}{\thesubsubsection)}{1em}{}

\titleformat{\paragraph}
{\normalfont\fontsize{12}{14}\bfseries  }{\thesection:}{1em}{}

 {
 \hypersetup{linkcolor=black}
 \tableofcontents
 }

\end{document}